# CRITICAL SUCCESS FACTORS FOR M-COMMERCE IN SAUDI ARABIA'S PRIVATE SECTOR: A MULTIPLE CASE STUDY ANALYSIS

Norah Suliman Alturaigi and Abdullah Abdulaziz Altameem

Department of Information System, Al-Imam Muhammad Ibn Saud Islamic University, Riyadh, KSA

*ABSTRACT*

*Many developing country firms are investing huge money in the sector of mobile commerce (m-commerce). Simplifying and understanding the factors which can impact on m-commerce success enables the organisations' managers to focus their efforts on the key areas of their m-commerce businesses, thereby contributing to the successful implementation of m-commerce. This study provides a clear understanding of m-commerce in the private sector in the Kingdom of Saudi Arabia and identifies the critical success factors of implementing m-commerce within the local business environment. A case study approach will be used for five Saudi companies, which use m-commerce, represented by: Alrajhi Bank, Souq.com, Saudi Electricity Company, Saudi telecom company (STC), and Saudi Airlines. This study represents a research in progress* and *interviews based on the literature to identify the key success factors for these companies in particular and in Saudi Arabia's private sector in general.*

## 1. INTRODUCTION

Many different companies are investing huge amounts of money in m-commerce applications and technologies that are used by the customers [1; 2]. Also, m-commerce is considered as a part of the development processes that have been conducted in the sector of e-commerce [3]. However, m-commerce has outperformed e-commerce. There is no need for a computer to purchase the services and goods one needs, meaning that consumers can purchase them from anywhere via their mobiles. M-commerce is an exchange of goods, services, and information that uses wireless hand-held devices like mobile phones [1; 2; 4]. A number of distinct features make m-commerce different from e-commerce. These include the use of product and service localization parameters, a more personalized shopping experience, ubiquity enhancement, the ability to instantly connect with consumers and more convenient shopping channels [1; 5]. M-commerce is driven by fundamental changes in the industries of information technology, telecommunications, media, and financial services [6]. With the use of mobile technology, m-commerce allows users to securely manage their businesses online, from any location and at any time.

The important factors of success for m-commerce implementation are thus some of the aspects of Saudi Arabia's business environment which impact the implementation and use of mobile commerce in the private sector. Based on the research problem area, there are two study questions that have been raised for this research:

DOI:10.5121/ijitcs.2015.5601                                                                                                                                   1

**RQ1:** What is the type and status of services provided though mobile commerce in Saudi Arabia's private sector?

**RQ2**: What are the critical success factors of implementing m-commerce in the private sector? From this question the following three sub-questions are raised:

- RQ2 A: What are the business factors that are critical to the success of m-commerce in the Saudi private sector?
- RQ2 B: What are the technical factors that are critical to the success of m-commerce in the Saudi private sector?
- RQ2 C: What are the acceptance factors that are critical to the success of m-commerce in the Saudi private sector?

In the theoretical sense, this research will fill the gap on the critical success factors of m-commerce compared to previous studies, by using a model that will combine the technical, business, and acceptance factors. In other words, this research will shed light on the critical factors, and their importance in a more comprehensive way.

### 1.1 Research Objectives

The main objectives of this study are:

- Know the current situation of m-commerce in the private sector in KSA.
- Determine the critical success factors are associated with the implementation of m-commerce in the private sector in Saudi Arabia.
- Give valued theoretical information to enhance m-commerce in KSA's private sector.
- Identify recommendations on the success factors of implementing m-commerce within the private sector to improve the decision-making process and enhance success.

### 1.2 Significance of the Study

The importance of this study comes from the significant role of mobile commerce applications in all aspects of life. M-commerce has various benefits. First, M-commerce applications impact process effectiveness and flexibility. They also lower operational costs, increase customer satisfaction, and improve the decision-making and productivity of organisations [7; 8]. It can also help managers make decisions anytime and anywhere, since it eliminates many time-consuming tasks [8; 9].

On the other hand, and on practical basis, this research will be significant as its results will help to improve the business environment in Saudi Arabia and increase its competitive advantage. This is because there are a number of development organizations that are investing considerable amounts of money in m-commerce developments. This research will also be beneficial in facilitating communication between employees, customers, and suppliers, by understanding their needs. Moreover, this research will introduce recommendations on the success factors of implementing m-commerce within the private sector.

## 2. LITERATURE REVIEW

Having a thorough understanding of the factors which can impact m-commerce will help the managers of these organisations focus their plans on the key areas of their m-commerce businesses, thereby contributing to the successful implementation of m-commerce [10].

As this research will be conducted in Saudi Arabia, it is worth starting by giving an idea and general background on information technology and particularly mobile commerce, in Saudi Arabia. In 2013, the Communication and Information Technology Commission (CITC) conducted research on the use of IT in Saudi Arabia, which began between 1990 and 2000 [11; 12]. It was during this time that IT realised unprecedented growth, with the economy of Saudi Arabia changing dramatically to support it. The government, businesses, and individuals used up a great deal of money on IT products and services; in 2013, around $6 billion had been spent on IT. In fact, during the economic crisis of 2012 and 2013, IT spending in Saudi Arabia was ongoing and trends were up. World Bank rankings had shown that Saudi Arabia plays main role in competitive market in the Middle East. According to the CITC, market spending grew by 17% in 2013. Regarding the local IT market in KSA, it is expected to spend $9.8 billion on IT. It is anticipated that the majority of this spend will be on hardware and IT services. As such, the growth of online retail in Saudi Arabia can be attributed to the development of new mobile applications that provide extensive business-to-business (B2B) and business-to-consumer (B2C) offerings. Several national banks, such as Al-Rajhi Bank, offer mobile services for banking and services based on online brokerage for the stock market as shown by [1].

Specifically, [13] defined the potential success factor as "*an aspect of a business environment that is influential in determining the outcome of a business activity*". Therefore, the factors of potential success are those aspects that will influence the implementation and the dependence on mobile commerce, thereby ensuring that the technology will be adopted by end-users. Hence, the suggested set of potential success technological factors of mobile commerce is found by checking the elements and factors that were impacted on e-business.

The research by [14] on the success of m-commerce proposed that a measure of the intent to adopt m-commerce could be used as an indirect method to evaluate its success, along with critical success factors. Critical success factors are discussed throughout previous studies from different point of views, and are now grouped into different categories that cover most of them. Those three categories of factors that are discussed in this research are: technological factors, business factors, and acceptance factors. Below is a detailed explanation of each category and its factors based on previous studies.

### 2.1 Technological Factors

The first factors are technological success factors. In order to be successful, m- commerce needs to incorporate the functionality and capability required to match traditional e-commerce offerings. On the one hand, because mobile commerce operates under additional restrictions, as mentioned above, there are extra and more particular perspectives that aid in its prosperity. However, on the other hand, these specific differences also result in a number of advantages that can aid its success [10].

The first of the technical restrictions is interface integrity. Desktop computers have a wide user base and are a typical aspect of everyday life. However, according to [15] the interactive and intuitive nature of mobile devices means that they are typically much easier to use than desktop computers. Moreover, the mobile professional is involved in several activities and is used in a dynamic environment. Also, [15] described how to make the most of the functionality that is available on mobile devices; applications that are specifically developed for m-commerce should ensure the design incorporates an effective user interface that is responsive and easy to use. On the other hand, complex menu structures and key sequences, installation scripts, and application errors and general protection faults should be avoided [16].

A further major limitation is security. Since security, undoubtedly, is of key consideration in the development of m-commerce applications. In a paper entitled "Is WAP ready for e-Commerce?" [17] Described how, while an extensive range of functionality can be incorporated into mobile devices, the use of such functionality is not without security implications. Also, [17] argued that the confidentiality of information is considered to be a number one priority, to reassure users that their communications are secured. It is important that any messages that are communicated via mobile devices are safe and that they are not intercepted by any unauthorized third parties. In addition to ensuring the safety of the data, it is also important to safeguard its integrity by ensuring that messages are not modified as they are sent from one server to another, from the sender to the recipient. A third priority relates to user authentication. Systems need to have the ability to ensure that user identities are authenticated and authorized appropriately.

It is also worth mentioning the TLS (Transport Layer Security) protocol, described by [18], which encrypts sensitive data and sends them over the web. WAP provides a secure network through integrating a wireless transport layer in the protocol stack. This encrypts all the data that are transmitted via the WAP link. [19], senior analyst at Ovum (Pty), argued that the current security functionality that is available in mobile devices is adequate for the exchange of small amounts of information. In the same year, [19] forecasted that future mobile devices would incorporate the functionality required to complete more data sensitive transactions such as online trading, banking, and payment services. Moreover, [20] supported the notion that to be suitable for performing high-value transactions, mobile devices needed to incorporate additional security features. Wireless transaction layer security offers one such method by which the WAP gateway can be secured, and online transactions that are completed using mobile devices can be enhanced to the required level [17].

Speed and efficiency represents a third limitation that is associated with the use of mobile devices, such as smartphones and hand-held computers, to complete online transactions. It is well known that, the operating environment incorporated in these devices is much more technically complicated than that employed in desktop PCs and laptops. As such, wireless devices typically have a limited battery life, low memory capacity, smaller screen displays and are slower to use for input purposes. Moreover, [21] claimed that these conditions will not make critical change in the near future, whereas, wireless phones are designed for compact, lightweight use. Furthermore, a more complicated communication environment is provided by wireless data networks compared to those of wired networks [22]. Three key factors that need to be taken into consideration when designing a wireless data network, which are latency, bandwidth, and the stability of the connection. It is expected that wireless networks to overcome these network limitations in the future, and will still deliver the necessary experience of user.

The fourth of the additional limitations is interoperability**.** According to [23], service providers have to feel safe that their investments are going to achieve advantages in the future. Thus, this is not able to be achieved except via tool and software provided by suppliers that is designed to work together. Service providers can select the tools and software they use from a variety of different vendors. It is critical that such providers choose suppliers who are WAP compliant and use solution parts that are aligned with the needs and functionality of their offering. Also, [23] claimed that one of the fundamental aspects of the WAP specification is that it incorporates the functionality needed to facilitate interoperability between varieties of WAP-compliant components.

The fifth of the additional limitations is the existence of user technology. This includes providing mobiles to users, and how much acquiring and using these devices costs. Such a factor may be inextricably linked with the ultimate success of m-commerce because the bigger the market of users, the higher the likelihood that electronic commerce will be successful is likely to be [22]. According to Global Magazine [16], the number of wireless subscribers has reached 200 million. However, users are more interested with the prices of handheld devices. According to [22], effective solutions provide great value at a low cost.

The sixth of the additional limitations is the availability of mobile infrastructure. [24] found that the current mobile architecture is designed for providers, developing software, and manufacturing mobiles. In order to deliver a sound and consistent infrastructure, there is a requirement to ensure that the solution that is employed is applied in a consistent manner across the various networks involved. As such, a single standard is required that performs effectively on all networks. This will help to ensure that all subscribers can access the benefits associated with each network. [24] Reinforced this point, highlighting the importance of the various manufacturers of the applicable devices utilizing the same software in their offerings to decrease development time and simplify the provision of support.

## 2.2 Business Factors

On the other hand, there are business success factors for mobile commerce. [25] Found that the critical business factors that donate to the success of m-commerce are associated with the causes that customers may or may not use mobile services. It is necessary considering besides to its recommendation corresponding for enhance these services. [26] Showed that one of the biggest advantages that is associated with the use of mobile services is the positive "price/service ratio" (which is the most popular response at 68%), the "comfort" of using it (the second most popular response at 55%), and the "independence of time and space" (seen as important by 48% of the respondents). On the contrary, it was found that the "high price of mobile access" is the critical reason for not using mobile services, with 61% of the respondents citing this reason. The second biggest reason attributed to people's decision not to use mobile services was poor service quality. (Cited by 56% of the respondents) and the third cause is "the lack of security" of mobile devices (cited by 53%). Finally, when ask the respondents to estimate the essential of the actions and procedures that must be considered by mobile operators to improve the mobile services they offer. Most respondents indicated that "lower price" is very important (61% cited this), while 39% of them suggested that "improved security" is important. A total of 34% of those surveyed disclosed that they perceived improvements to the device comfort to be important; as such, this factor ranked as the third most important dimension.

According to [27], a generic m-commerce strategy should be proposed that meets the adoption goal. It is important to measure the objectives of marketing at the outset. For example, one might aim to increase the m-commerce consumer adoption rates from a rate of 5% to 25%. The resulting strategy is the "penetration pricing strategy". [28]; [29] argued that applying the policy of pricing in the sector of industry and commerce seems to provide a chance to challenge other businesses. Strategies include charging for a product or service with a low price, in line with the target to accelerate its use and, subsequently, increase market share. It is crucial to mention that any pricing strategy is important in price-sensitive markets. The survey results for the survey that was conducted by [30] showed that in Europe, the high cost of m-commerce services form the most essential cause to avoid using these services. Moreover, it was finished that investment in various technological offerings include developing improvements to the devices that are available, the security incorporated in them, the bandwidth they offer and the design of the shopping interfaces and applications that are available, which provides another chance to develop m-commerce services and achieving customer satisfaction.

Furthermore, [31] argued that elements of marketing are effective in manipulating the success of a product, which is the device in this study. The "promotion" (e.g., communicating improvements in security) besides with the actual manipulation of the "price" part has an energetic role in the success of m-commerce. Furthermore, According to [15], in terms of mobile devices, "good coverage" is interchangeable with the traditional notion of the "place" factor of the marketing mix. As such, the extent to which a customer can gain network coverage is an essential factor in their selection of a mobile operator.

[32] Pointed out that businesses should decrease the price of mobile access and develop the service quality and customer support they provide. But then, they argued that efforts must take place on designing more suitable mobile devices like large screens, in addition to producing shopping applications and user interfaces that are intuitive and simple to use. The challenge of bandwidth and security must also be fixed and this type of "solution" must be contacted via exact promotional campaigns to present and potential users. Finally, [32] argued that more emphasis must take place on a coverage strategy, as it was proven to form a common part of mobile operator selection criterion.

Both [33] and [34] applied the diffusion of innovation theory within their respective research studies. According to their findings, the mobile phone users who had utilized m-commerce services at that time represented "innovators"; those innovators show interest to issues like pricing and promotion in their adoption of m-commerce. However, [35] found that at different times, Individuals who are quick to use innovative offerings are regarded as "early adopters." Mobile operators must be aware of these individuals. Consumers are more heavily involved in the local community than innovators and typically interact with those selling new products on a regular basis. As opinion leaders, they have the ability to influence the opinions and purchase choices of other consumers. In addition, [36] argued that business plans like reducing price, improvement of customer service and etc. and technological yields, such as the improvement of mobile devices, resolution of security problems, etc. represent important aspects that must be delivered to increase use of, and openness to, m-commerce services.

## 2.3 Acceptance Factors

Thirdly, there are the acceptance success factors for m-commerce, in reference to the theory of information systems success recommended by [37]. This theory takes into consideration three primary success factors that are required in any information system: system quality, information quality, and service quality, since these factors significantly influence user consumption and usage. Consequently, Theories that seek to identify what factors contribute to the success of information systems have gained increasing attention, and more and more researchers are applying the available models in the context of e-commerce [38; 39]. [40]; [41] explained the features that emerging information technology contributes to the success of mobile sites. These characteristics allow users to reach sites of mobile and gain information from anywhere and at any time.

The information systems success theory, the model of technology acceptance (TAM) and the theory of trust can be used in conjunction to clarify the levels of user adoption of m-commerce. TAM provides useful insights into the perceived ease of use and usefulness of an application to allow researchers to understand these elements and the extent to which these have an impact on a user's willingness to use the information technology available. Identifying user opinions of perceived ease of use and perceived usefulness is deemed to be of significance because consumer acceptance of novel information technologies are widely regarded as being influenced by these two factors [40; 41]. Perceived usefulness relates to the extent to which a user believes the technology will provide them with an efficient and effective service, while perceived ease of use reflects individual's views of how difficult it is to use the technology to achieve the given objective. According to [15], user's perceptions of the ease of use and usefulness of mobile sites will have a direct impact on their evaluation of the quality of the system, information, and service that is available.

Moreover, [42] mentioned that, customers need to be confident that they can easily identify risks and find the mobile sites usages are simple. The quality of the information provided on a system and the quality of the application itself directly impacts user's perceptions of the extent to which the mobile offering can be trusted. It is essential that mobile service providers develop high-quality applications that contain useful and reliable information. The TAM theory can be applied to develop useful insights into the perceived ease of use and usefulness of an application to allow researchers and marketers to develop a deeper understanding of what factors contribute to a consumer's acceptance of a mobile site.

The TAM model is also common used to explain behaviour of mobile adoption. [43]; [44]; [45] studied the user adoption rates of mobile data offerings. Their findings indicated that a range of different factors can influence the extent to which a user is willing to embrace new technology and, thus, the success of new offerings. These include users' perceptions of cost/benefit, fun, ease of use, entertainment, and uniqueness. Specifically focusing on mobile payment systems, according to [46], the perceived ease of use and usefulness of an application that allows mobile payments correlates with a user's willingness to use it. [42] Named a number of factors that can influence whether or not a user adopts mobile Internet services. These included the variety of services available, the quality of access, cost, and perceived ease of use. Research by [47] provided a different perspective. They found that lack of trust will have a negative impact on a user's wiliness to adopt the use of a technology, and they argued that this can be further undermined by views that a competitor's future offering will be superior. The virtual and anonymous nature of online transactions entails that trust is of extreme importance and, as such, many researchers in the online commerce context have focused on the methods of achieving and sustaining the trust required. [48] Presented a model in which the TAM approach was combined with trust indicators to better understand online purchase behaviour. The results of the study revealed that a user's perceived ease of use impacted the extent to which they trusted the application and believed it to be useful.

Furthermore, [49] developed a scale that they employed to measure user's initial perceptions of trust when they encountered a mobile application. They found that the quality of a website has a direct impact on these initial notions of trust. Further work by [50], found that technological perceptions and perceived ease of use impacted user's initial trust in a mobile application. [51] Performed a review of which factors determine the extent to which a user trusts an online resource. Their findings revealed that both perceived usefulness and ease of use can have an impact on this sense of trust, and that both these factors will subsequently influence the rate at which an application is adopted and is ultimately successful. M-commerce based on internet has great risk and also uncertainty. As an example, it is known that it is easy to hackers to attack mobile networks.

In other words, [52] argued that the seen usefulness and seen ease of use plays main role on end-users aim to use m-commerce. Moreover, cost and subjective norms are critical antecedents of end-users' intention. For instance, in the USA, privacy of consumer, innovativeness, seen usefulness, seen enjoyment, and compatibility are shown as factors that influence intentions to use m-commerce. It has also been argued that many of these distinguishes can be indicated by factors of culture and economy. Finally, [53] argued that the way in which business, technology, and acceptance factors interact indicates that further research across a range of different disciplines is required to ultimately improve the offering.

International Journal of Information Technology Convergence and Services (IJITCS) Vol.5, No.6, December 2015## 3. METHODOLOGY

In this study, the design will be descriptive, the researcher will adopt interviews as a tool to answer the research questions and achieve the main objectives of the study. This research uses the qualitative approach to measures the critical success factors of implementing m-commerce in local businesses. In this research, the case study will be consist of interviews in five private sector companies in Saudi Arabia represented by (Alrajhi Bank, Souq.com, Saudi Electricity Company, Saudi Telecom Company (STC), and Saudi Airlines), However, it is believed that the case study may be more 'relatable' to other companies in the same fields [54].

## 4. CONCLUSION

The primary purpose of this research is to determine the critical success factors of implementing m-commerce in the private sector in Saudi Arabia. Also, represents a research in progress, whereas the interviews will be conducted on companies that have already applied m-commerce in KSA. Thus, five main local business companies were chosen, including Al Rajhi Bank, Saudi Airlines, Saudi Electricity Company, Souq.com and Saudi Telecom. All these companies implemented m-commerce within their local business environment. Through the research, it will clarify the findings of previous studies which concerned on the critical success factors of m-commerce. Also, aim to distinguish it from previous studies and address the current gaps in the literature.

## 5. PROGRESS TO DATE AND NEXT STEPS

This research represents a research in progress and the researcher works in interviews and collecting data from five companies (Alrajhi Bank, Souq.com, Saudi Electricity Company, Saudi telecom company (STC), and Saudi Airlines).

## 6. REFERENCE

[1] Al-Muhtib, I. M. (2011) M-commerce in Saudi Arabia. M.S. thesis: University of Heriot Watt, On, U.K.
[2] Tiwari, R., Buse, S. and Herstatt, C. (2006) From electronic to mobile commerce: Opportunities through technology convergence for business services. Asia Pacific Tech Monitor, 23(5), 38-45.
[3] GS1 Mobile Com. (2008) Mobile Commerce : opportunities and challenges, A GS1 Mobile Com White Paper, Available At: http://www.gs1.org/docs/mobile/GS1_Mobile_Com_Whitepaper.pdf
[4] Ding, M. S. and Unnithan, C. (2002) Mobile commerce (mCommerce) security. An appraisal of current issues and trends. In 12th Annual BIT Conference Proceedings. MMUBS Manchester Metropolitan University Business School, p. 19.
[5] Ngai, E. W. and Gunasekaran, A. (2007) A review for mobile commerce research and applications. Decision Support Systems, 43(1), 3-15.
[6] Hsieh, C. T. (2007) Mobile Commerce: Assessing New Business Opportunities. Communications of the IIMA, 7(1), 87-100.
[7] Wang, H. and Xu, Q. (2012) Improving M-commerce through Enterprise Mobility. In Management of e-Commerce and e-Government (ICMeCG), 2012 International Conference on (pp. 211-215). IEEE.
[8] Gebauer, J. and Shaw, M. J. (2004) Success factors and impacts of mobile business applications: results from a mobile e-procurement Study. International Journal of Electronic Commerce, 8(3), 19-41.
[9] ii, I. C. (2008) Emerging Value Propositions for M-commerce, Journal of
[10] Wei, T. T., Marthandan, G., Chong, A. Y. L. and Ooi, K. B. (2009) what drives Malaysian m-commerce adoption? An empirical analysis. Industrial Management & Data Systems, 109(3), pp. 370-88.8

International Journal of Information Technology Convergence and Services (IJITCS) Vol.5, No.6, December 2015[36] Park, K. S., Koh, C. E. and Nam, K. (2010) Perceptions of RFID technology: a cross-national study, Industrial Management & Data Systems, 110(5), pp. 682-700.

[37] Pavlou, P. A. and El Sawy, O. A. (2010) the 'third hand': IT-enabled competitive advantage in turbulence through improvisational capabilities, Information System Research, 21(3), pp. 443-71

[38] Trainor, K., Rapp, A., Skinner, L. and Schillewaert, N. (2011) Integrating information technology and marketing: an examination of the drivers and outcomes of e-marketing capability, Industrial Marketing Management, 4, pp. 162-74.

[39] Shankar, V., Venkatesh, A., Hofacker, C. and Naik, P. (2010) Mobile marketing in the retailing environment: current insights and future research avenues, Journal of Interactive Marketing, 24, pp. 111-20.

[40] Gable, G. G., Sedera, D. and Chan, T. (2010) Re-conceptualizing information system success: the IS-impact measurement model. Journal of the Association for Information Systems, 9(7), pp. 377-408.

[41] Cheung, C. M. K. and Lee, M. K. O. (2009) User satisfaction with an internet-based portal: an asymmetric and nonlinear approach. Journal of the American Society for Information Science and Technology, 60(1), pp. 111-22.

[42] Shin, Y. M., Lee, S. C., Shin, B. and Lee, H. G. (2010). Examining influencing factors of post-adoption usage of mobile internet: focus on the user perception of supplier-side attributes. Information Systems Frontier, 12(4).

[43] Lu, J. (2013) Are personal innovativeness and social influence critical toMcontinue with mobile commerce? Internet Research, 24(2), pp. 134-159, Emerald Group Publishing Limited.

[44] Nor, K. and Jahanshahi, A. (2011) From Mobile to Mobile Commerce: An Overview in the Indian Perspective - 2nd International Conference on Business and Economic Research, 35(8), pp. 982-1003.

[45] Wu, X., Chen, Q., Zhou, W. and Guo, J. (2010) A review of mobile commerce consumers' behavior research: consumer acceptance, loyalty and continuance (2000-2009), International Journal of Mobile Communications, 8(5), pp. 528-60.

[46] Kim, C., Mirusmonov, M. and Lee, I. (2010) An empirical examination of factors influencing the intention to use mobile payment. Computers in Human Behavior, 26(3), pp. 310-22.

[47] Consulting, C. (2011) Consumer Market Study on the Functioning of E-Commerce and Internet Marketing and Selling Techniques in the Retail of Goods, Final Report Part 1: Synthesis Report, prepared for the Executive Agency for Health and Consumers on behalf of the European Commission.

[48] Akter, S., D'Ambra, J. and Ray, P. (2011) Trustworthiness in m-health information services: an assessment of a hierarchical model with mediating and moderating effects using partial least squares (PLS). Journal of the American Society for Information Science & Technology, 62(1), pp. 100-116.

[49] McKnight, D. H., Choudhury, V. and Kacmar, C. (2012) Developing and validating trust measures for e-commerce: an integrative typology. Information Systems Research, 13(3), pp. 334-59.

[50] Benamati, J. S., Fuller, M. A., Serva, M. A. and Baroudi, J. A. (2010) Clarifying the integration of trust and TAM in e-commerce environments: implications for systems design and management. IEEE Transactions on Engineering Management, 57(3), pp. 380-93.

[51] Beldad, A., de Jong, M. and Steehouder, M. (2010) How shall I trust the faceless and the intangible? A literature review on the antecedents of online trust.

[52] Dai, H. and Palvia, P. (2009) Mobile Commerce Adoption in China and the United States: A Cross-Cultural Study. The DATA BASE for Advances in Information Systems. 40(4). Made available courtesy of The ACM Special Interest Group on Management Information Systems (SIGMIS) http://dl.acm.org/citation.cfm?doid=1644953.1644958 .

[53] Bianchi, C. and Andrews, L. (2012) Risk, trust, and consumer online purchasing behaviour: a Chilean perspective. International Marketing Review, 29(3), pp. 253-275.

[54] Denscombe, M. (2003) The Good Research Guide for Small Scale Social Research Projects. (2ed Ed). Open University Press: Philadelphia.
10